\documentclass[aps,prl,superscriptaddress,showkeys,preprintnumbers,showpacs,longbibliography,nofootinbib,floatfix,twocolumn]{revtex4-2}
\usepackage[utf8]{inputenc}
\usepackage{amsmath}
\usepackage{amsfonts}
\usepackage{amsthm}
\usepackage{amssymb}
\usepackage{float}
\usepackage{braket}
\usepackage{graphicx}
\usepackage{xcolor}
\usepackage{url}
\usepackage[colorlinks=true, pdfstartview=FitV, linkcolor=red, citecolor=blue, urlcolor=blue]{hyperref}
\usepackage{slashed}
\usepackage[normalem]{ulem}

\graphicspath{{./Figures/}}

\newcommand{\be}{\begin{equation}}
\newcommand{\ee}{\end{equation}}
\newcommand{\beq}{\begin{equation}}
\newcommand{\eeq}{\end{equation}}
\newcommand{\bea}{\begin{eqnarray}}
\newcommand{\eea}{\end{eqnarray}}

\newcommand{\tr}{\mathrm{tr}}

\newcommand{\calT}{\mathcal{T}}

\newcommand{\Tr}{\mathrm{Tr}}

\newcommand{\newtext}[1]{{\color{black}{#1}}}

\newcommand{\nn}{\nonumber \\}

\begin{document}
\title{Thermal modifications of mesons and energy-energy correlators \\ 
from real-time simulations of a $U(1)$ lattice gauge theory}

\preprint{CERN-TH-2025-133}

\author{João Barata}
\affiliation{CERN, Theoretical Physics Department, CH-1211 Geneva 23, Switzerland}

\author{David Frenklakh}
\email[]{dfrenklak@bnl.gov}
\affiliation{Department of Physics, Brookhaven National Laboratory, Upton, New York 11973-5000, USA}

\author{Swagato Mukherjee}
\affiliation{Department of Physics, Brookhaven National Laboratory, Upton, New York 11973-5000, USA}

\bibliographystyle{unsrt}

\begin{abstract}
We investigate 
thermal properties of a $U(1)$ lattice gauge theory 
in $1+1$-dimensions through real-time simulations. We extract the spectral functions directly coupling to the pseudoscalar and scalar mesons, demonstrating the thermal modifications of these states with increasing 
temperatures. Introducing the notion of energy-flow operators, we quantify the temporal build-up of correlations in the energy flows across the lattice. We demonstrate that energy-energy correlators fail to factorize to products of energy flows, both in the vacuum and at 
nonzero-temperature, indicating the presence of non-trivial correlations in the quantum states. Our results constitute a first real-time \textit{ab-initio} study of bound state thermal broadening and finite temperature energy-flow correlations in a gauge theory, providing a benchmark for future studies of hadronic matter under extreme conditions.

\end{abstract}

\maketitle



\emph{\textbf{Introduction.}}  Characterizing the properties and dynamics of quantum systems under extreme conditions remains one of the central open theoretical problems across physics, from condensed-matter~\cite{Bloch_2008,Mitra_2018,Bernard:2016nci} to high-energy theory~\cite{Busza:2018rrf, Berges:2020fwq}. A deeper understanding of these regimes is not only crucial for unraveling the inner workings of quantum field theories, but also essential to describe a myriad of phenomena observed in Nature across all scales, from the formation of the early Universe~\cite{Turner:2022gvw} to the dynamics of atomic systems~\cite{ludlow2015opticalatomicclocks}.

In high-energy nuclear physics, elucidating the behavior of quantum chromodynamics (QCD) at finite temperature and baryon density is essential for explaining how matter forms and evolves within the Standard Model. Over the past five decades a succession of high energy nuclear experiments --- culminating in RHIC and LHC's ultrarelativistic heavy ion physics programs~\cite{STAR:2005gfr, ALICE:2010suc} --- has provided an unprecedented opportunity to test the properties of hot and dense QCD, including its far-from-equilibrium dynamics. These efforts led to the first experimental observation of a quark–gluon plasma state~\cite{Heinz:2000bk}, which filled the primordial Universe in the first few microseconds after the Big Bang.

Despite the numerous theoretical and experimental advances motivated by these physics programs, fundamental questions about the properties of QCD matter under extreme conditions remain open. In this respect one could argue that two theoretical limitations stand out. First, although finite temperature ($T$) equilibrium (Euclidian) lattice QCD calculations have allowed precise mapping of several properties of the theory, see e.g.~\cite{Ratti:2018ksb}, notorious sign problems prevent the access to the real-time dynamics and systems at finite density, needed to describe the space-time evolution of matter and the phase structure of the theory~\cite{Nagata:2021ugx}. Second, and as a result of the previous point, such dynamical observables are thus studied through effective descriptions, e.g. QCD effective kinetic theory~\cite{Arnold:2002zm} or relativistic viscous hydrodynamics~\cite{Romatschke:2017ejr}. In practice, the respective weak-coupling and near-equilibrium assumptions underlying these pictures are typically pushed beyond their theoretical applicability, preventing a true first-principle description of the observed phenomena. Moreover, how these different effective descriptions can be properly combined is still an open question. 

Therefore, bridging this gap requires novel theory tools capable of following real-time dynamics of QCD at finite $T$ from a first-principle formulation. In this regard, emerging technologies and methods from Quantum Information Science (QIS) --- such as digital quantum computers or analogue quantum simulators --- promise exactly the capability of tracking the real-time dynamics of quantum systems from first principles, sidestepping the sign problem, see e.g.~\cite{Martinez:2016yna,Wiese:2013uua}. Extending these QIS techniques to gauge theories at finite temperature could, one day, allow for an entirely new way to describe and understand the complex dynamics of heavy ion collisions and ultimately the properties of QCD matter under extreme conditions.

In this work, we take a first step in this direction by performing real-time simulations in Schwinger model, a $U(1)$ gauge theory in $1+1$ dimensions, at finite $T$. This extends a large range of recent works on this model at $T=0$, see e.g.~\cite{Banuls:2019bmf,Banuls:2019rao} for recent reviews. We present results for the temperature dependent spectral functions of the theory’s lightest mesonic excitations, revealing how the inter-fermion potential is thermally screened by the hot medium. We directly observe that the mass of the bound states 
remain largely 
unchanged at finite $T$, indicating that the real part of the respective inter-fermionic potential is unmodified. Conversely, the width of these states grows with temperature, indicating the rise of the imaginary part of the potential similarly to what is seen in lattice QCD studies. The pattern of sequential \textit{melting}, paradigmatic as \textit{thermometer} of the bulk matter produced in heavy-ion collisions~\cite{Rothkopf:2019ipj, Matsui:1986dk}, is not cleanly observed although there is distinction between the studied bound states. We complement the discussion by studying the behavior of detector observables at finite $T$, computing energy-correlation functions that play the role of calorimeter cells and track how short-range correlations propagate to asymptotic distances, see e.g.~\cite{Moult:2025nhu} for a recent review. We find that at finite $T$ more non-trivial correlations develop compared to the $T=0$ scenario. Technical aspects of our calculations and details of the lattice model considered are introduced beforehand.

\vspace{0.3cm}

\emph{\textbf{Lattice Schwinger model at finite $T$.}} 
The Schwinger model corresponds to a $U(1)$ gauge quantum field theory in $1+1$ spacetime dimensions~\cite{Schwinger:1951nm}, analog to the QED in $3+1$-d. This theory describes the interaction of a single-flavor fermion $\psi$, with mass $m_c$, with the gauge field $A^\mu$. The matter-gauge coupling is $g_c$; \newtext{the subscript ``c" in the model parameters stands for \textit{continuum}.} While this theory is in many ways simpler than QCD in $3+1$-d, they share several key qualitative features. In particular, among other features, the Schwinger model possesses an inter-fermionic confining potential, spontaneous chiral symmetry breaking, and chiral anomaly; see e.g.~\cite{Coleman:1976uz, Schwinger:1951nm, Mandelstam:1975hb, Coleman:1975pw, Banks:1975gq}. Unlike higher dimensional theories, the Schwinger model is dual to a bosonic interacting theory~\cite{Coleman:1976uz}, where the fundamental particle is the so called Schwinger boson. In the $m_c=0$ limit, this massive boson, with mass  $g_c/\sqrt{\pi}$, becomes non-interacting and the theory can be solved analytically through Abelian bosonization methods.

Although for general $m_c/g_c$ the theory can not be solved analytically, its real-time properties can be numerically studied using lattice simulations in the Hamiltonian picture, see e.g.~\cite{Wiese:2013uua,Banuls:2019bmf}. Recently, such numerical approaches have become practically feasible, with the progress of new quantum technologies and related tools. Currently they are particularly suited for the simulation of lower dimensional theories.\footnote{See e.g.~\cite{Gonzalez-Cuadra:2024xul,DiMarcantonio:2025cmf,Felser:2019xyv} for recent studies of higher dimensional theories.} 

In the continuum, the Schwinger model in the temporal gauge $A_0 = 0$ is described by the Hamiltonian:
\begin{equation}
   \hspace{-.4 cm} H_{\rm cont} =\int dx \left [ \frac{1}{2}F_{01}^2+\bar{\psi}(-i\gamma^1\partial_1+g_c\gamma^1A_1+m_c)\psi \right ],
\end{equation}
where $F_{01} = \partial_0A_1$. It can be mapped to a Kogut-Susskind lattice Hamiltonian~\cite{Kogut:1974ag,Hamer:1997dx}:
\begin{align}
    H_{\rm lat} &= -\frac{i}{2a}\sum_{n=1}^{N-1}
\big(U_n^\dag\chi^\dag_{n}\chi_{n+1}-U_n\chi^\dag_{n+1}\chi_{n}\big) \nonumber \\
    &+ m\sum_{n=1}^N (-1)^i\chi_n^\dag\chi_n 
    + \frac{ag^2}{2}\sum_{n=1}^{N-1}L_n^2 \, , \label{eq:Ham_lat}
\end{align}
where we introduced staggered fermionic variables $\chi_n$, gauge link operators $U_n = e^{-iagA_1(an)}$ and electric field strength operators $L_n= E(an)/g$. The system has volume $Na$ given in terms of lattice spacing $a$ and the total number of sites $N$. We use $m$ and $g$ to denote the mass and coupling in the lattice model, respectively. We further employ the finite lattice spacing correction to the mass, $m \to m-g^2a/8$, introduced in~\cite{Dempsey:2022nys} to preserve chiral symmetry in the massless limit.

The Hamiltonian in Eq.~\eqref{eq:Ham_lat} possesses a residual gauge freedom, namely $\chi_n\rightarrow\Omega_n \chi_n, U_n\rightarrow \Omega_{n+1}^\dagger U_n\Omega_n$. Performing a gauge transformation with $\Omega_n = \prod_{i=1}^{n-1} U_i$, we eliminate the link operators from the Hamiltonian. Furthermore, the electric field operators must obey the discretized version of Gauss' law:
\begin{equation}
    L_n - L_{n-1} = Q_n ~,
\end{equation}
where $Q_n = \chi_n^\dagger\chi_n + \frac{(-1)^n - 1}{2}$ is the electric charge operator. With open boundary conditions, no background electric field, and in the total charge zero sector, Gauss' law can be solved as $L_n = \sum_{i=1}^n Q_i$. Thus the system is described exclusively in terms of fermionic variables with a finite-dimensional Hilbert space, at the expense of introducing an all-to-all interaction in the electric field energy.

To make  contact with the language of QIS, fermionic degrees of freedom can be mapped to spins by the means of the Jordan-Wigner transformation~\cite{Susskind:1976jm}:
\begin{equation}
    \chi_n = \frac{\left(X_n-iY_n\right)}{2}\prod_{k=1}^{n-1} (-i Z_k) \,,
\end{equation}
where $X_i,Y_i,Z_i$ are the Pauli matrices acting in the spin-1/2 Hilbert space of site $i$. The respective lattice Hamiltonian reads:
\begin{align}
    H &= \frac{1}{4a}\sum_{i=1}^{N-1} \left(X_iX_{i+1} + Y_iY_{i+1}\right) + \frac{m}{2}\sum_{i=1}^N (-1)^i Z_i \nonumber \\
    &+ \frac{ag^2}{2}\sum_{i=1}^{N-1}\bigg(\sum_{j=1}^{i} \frac{Z_j+(-1)^j}{2}\bigg)^2\,.
\end{align}

In this work we explore properties of the theory at finite $T$. In general, a finite temperature state is a (statistically) mixed quantum state described by the thermal density matrix 
\begin{equation}
    \rho_\beta = \frac{e^{-\beta H}}{\Tr\, e^{-\beta H}} \,,
\end{equation}
where $\beta=T^{-1}$ is the inverse temperature. Thermal expectation value of any operator $\cal O$ is given by
\begin{equation} \label{eq:thermal_rho_O}
    \langle {\cal O}\rangle_\beta = \Tr (\rho_\beta {\cal O})\,.
\end{equation}
Due to the presence of a conserved charge --- the total electric charge of the system --- the thermal density matrix is block-diagonal in the basis where the charge operator is diagonal, for instance in the computational basis where the Pauli operators take their familiar form. Furthermore, since the electric charge is confined in the Schwinger model, non-zero charge sectors decouple in the infinite volume limit. Thus the charge zero sector is of main interest, and in the following we will always restrict ourselves to the chargeless sector. Any operators we consider also commute with charge, and Eq.~\eqref{eq:thermal_rho_O} holds if $\rho_\beta$ and $\cal O$ are understood as their respective charge zero blocks.

For sufficiently small systems it is possible to perform a full diagonalization of the Hamiltonian, i.e. find the eigenstates $|n\rangle$ and the spectrum $E_n$ such that
\begin{equation} \label{eq:eigenstates}
    H|n\rangle = E_n|n\rangle \,,
\end{equation}
so that the thermal expectation value of an operator $\cal O$ is
\begin{equation}
    \langle{\cal O} \rangle_\beta = \sum_n p_n \langle n|{\cal O}|n\rangle\,,
\end{equation}
where $p_n = e^{-\beta E_n}/Z$, with $Z=\sum_n e^{-\beta E_n}$, are the Boltzmann weights. As noted above, sums over $n$ include only the eigenvalues with zero charge. The dimension of the corresponding Hilbert space is $D = \begin{pmatrix} N/2\\N\end{pmatrix}$ for a system with $N$ lattice sites. All the results reported below are obtained with this method of exact diagonalization.

\vspace{0.3cm}

\emph{\textbf{Spectral functions and thermal modifications of mesons.}}
Spectral functions are an indispensable tool for understanding the properties of the excitation spectrum in an interacting quantum theory; in QCD they play a critical role in e.g. describing the bulk matter properties in heavy ion experiments. In particular, they have been used to study quarkonium melting~\cite{Rothkopf:2019ipj, Brambilla:2010vq,Brambilla:2016wgg}, dilepton and photon emission rates~\cite{Wu:2024pba,Garcia-Montero:2024lbl,Garcia-Montero:2023lrd,David:2019wpt,Turbide:2003si}, and transport coefficients~\cite{Kaczmarek:2022ffn}. Since a spectral function is related to a real-time two-point correlator of the meson-like operators, it is most natural to study it with Hamiltonian methods. This has previously been performed in a non-gauge theory, see~\cite{Banuls:2022iwk}. In what follows we discuss the spectral properties of such real-time two-point correlation functions in the lattice Schwinger model at finite $T$.

Consider a time-ordered two-point function of an arbitrary local operator $\cal O$: 
\begin{align}
    D_{\cal O}(x,t,x_0,t_0) \equiv \langle{\cal T}{\cal O}(x,t) {\cal O}^\dagger(x_0,t_0)\rangle_\beta \, , \label{eq:2pt}   
\end{align}
where in the context of QCD, the operator $\cal O$ is often chosen as a meson operator, $M(x,t) = \bar q(x,t)\Gamma q(x,t)$, with $\Gamma$ denoting an appropriate combination of Dirac matrices corresponding to the meson's quantum numbers. The correlation function can be decomposed into commutator and anticommutator parts, reflecting the statistical function $F$ and the spectral function $\rho$~\cite{Rothkopf:2019ipj}:
\begin{align}
     D_{\cal O}(x,t,x_0,t_0) &=  F_{\cal O}(x,t,x_0,t_0) \nn 
     &- \frac{i}{2}{\rm sign}(t-t_0)\rho_{\cal O}(x,t,x_0,t_0) \, .
\end{align}

In the Schwinger model, there are several species of mesons, with the lowest lying one being the pseudoscalar state. In the massless fermion case, as noted above, the Schwinger model is exactly solvable via bosonization, with the free Schwinger boson being precisely this pseudoscalar meson. Naturally, the two-point function of axial charge operators, $q_5 (x)\equiv \bar\psi(x)\gamma^0\gamma^5\psi(x)$, receives contributions from this pseudoscalar state.\footnote{Note that, due to the Dirac matrix algebra in $1+1$ dimensions, $q_5(x) = j(x) = \bar \psi(x)\gamma^1\psi(x)$, so one can identify the pseudoscalar and the vector meson state, as sometimes found in the literature.} We thus consider the two-point correlator:
\begin{equation}
    D_p(x,t) = \langle {\cal T} q_5(x,t) q_5(0,0)\rangle_\beta \, ,
\end{equation}
where the subscript $p$ stands for pseudoscalar. In momentum space, focusing on the zero spatial momentum mode, we find 
\begin{equation}
    D_p(\omega) =  \frac{1}{V}\lim_{\epsilon\rightarrow 0} 
 \int_{-\infty}^\infty dt\,e^{i\omega t}\langle {\calT} Q_5(t)Q_5(0)\rangle_\beta e^{-\epsilon|t|}     \,,
\end{equation}
where we suppressed $k=0$ in the notation; $V$ is the 1-dimensional spatial volume of the system and $Q_5$ is the global axial charge:  $Q_5(t) = \int_0^L dx \, q_5(x,t)$. To target the second excited state, corresponding to the a scalar meson at strong coupling~\cite{Coleman:1976uz,Adam:2002pd}, we further consider the higher weight correlator:
\begin{equation}
    D_s(\omega) = \frac{1}{V}\lim_{\epsilon\rightarrow 0} 
 \int_{-\infty}^\infty dt\,e^{i\omega t}\langle {\calT} Q_5^2(t)Q_5^2(0)\rangle_\beta e^{-\epsilon|t|}     \, .
\end{equation}
This structure is motivated by the strong coupling limit where one can understand the scalar state as a bound state of two pseudoscalar mesons. Indeed, in this limit $m_S = 2 m_P + \mathcal{O}(mg^{-1})$~\cite{Adam:2002pd}, where $m_S$ and $m_P$ are the scalar and pseudoscalar meson masses, respectively. 

The procedure to extract either of these correlators is similar; we focus the discussion now on the pseudoscalar sector. We define the finite $T$ correlator
\begin{equation}
    C_{Q_5}(t) = \langle {\cal T} Q_5(t) Q_5(0) \rangle_\beta \equiv \tr(\rho_\beta Q_5(t) Q_5(0)) \, ,
\end{equation}
where $Q_5(t)$ evolves according to the Heisenberg picture
\begin{equation}
    Q_5(t) = e^{iHt} Q_5(0) e^{-iHt}\,.
\end{equation}
The discretized version of the global axial charge operator is
\begin{equation}
    Q_5(0) = \sum_n \frac{Y_nX_{n+1}-X_nY_{n+1}}{4}\, ,
\end{equation}
after applying the Jordan-Wigner representation of the fermionic fields. Defining the matrix elements of the axial charge operator:
\begin{equation}
    c_{mn} = \langle m |Q_5(0)|n\rangle \, ,
\end{equation}
where $\{|n\rangle\}$ are the Hamiltonian eigenstates in Eq.~\eqref{eq:eigenstates},
one finds an exact expression for the time-dependence of the axial charge correlator
\begin{equation}
   {\hspace{-.5 cm}} \langle {\calT} Q_5(t)Q_5(0)\rangle_\beta = \frac{1}{Z}\sum_{n,m} |c_{nm}|^2 e^{-\beta E_n} e^{i(E_n-E_m)|t|} \, ,
\end{equation}
where at zero temperature the double sum is replaced by a single sum, setting $n=0$. Furthermore, the frequency domain expression for the spectral function can also be determined analytically~\cite{Rothkopf:2019ipj}:
\begin{align}
    &\rho(\omega) = \sum_{n,m} |c_{nm}|^2  p_n [\delta(\Delta E_{nm}+\omega) - \delta(\Delta E_{mn}+\omega)] \nonumber
    \\& {\hspace{-.4 cm}} = \sum_{n,m} |c_{nm}|^2  p_n \frac{4\epsilon \omega \Delta E_{mn}}{(\Delta E_{nm}^2-\omega^2)^2  + 2\epsilon^2 (\Delta E_{nm}^2+\omega^2)+\epsilon^4}\bigg|_{\epsilon\rightarrow 0} , \label{eq:freq_current_correlator} 
\end{align}
where we defined energy differences $\Delta E_{nm}\equiv E_n-E_m$ and in the second line introduced a regulator $\epsilon$ for convergence. From a practical real-time simulation viewpoint, the $\epsilon$ regulator suppresses the contributions from $|t|\gtrsim1/\epsilon$ reflecting the finite time intervals accessible in a simulation. Note that with exact diagonalization, one can directly compute $c_{nm}$ and $\Delta E_{nm}$, so performing a real-time evolution numerically is not necessary. However, for a simulation with quantum hardware or with tensor network methods an actual time evolution would be required. Our early attempts using the latter revealed a rapid increase of the bond dimension of the matrix product state during the real-time evolution of a thermal state in the purification approach~\cite{Verstraete:2004gdw, Zwolak:2004nwu,Feiguin:2005jud}, even for modest ($N\sim 40$) system sizes. Thus we leave a detailed tensor network study to the future, and focus on exact diagonalization in this work.

\begin{figure}
    \centering
\includegraphics[width=\linewidth]{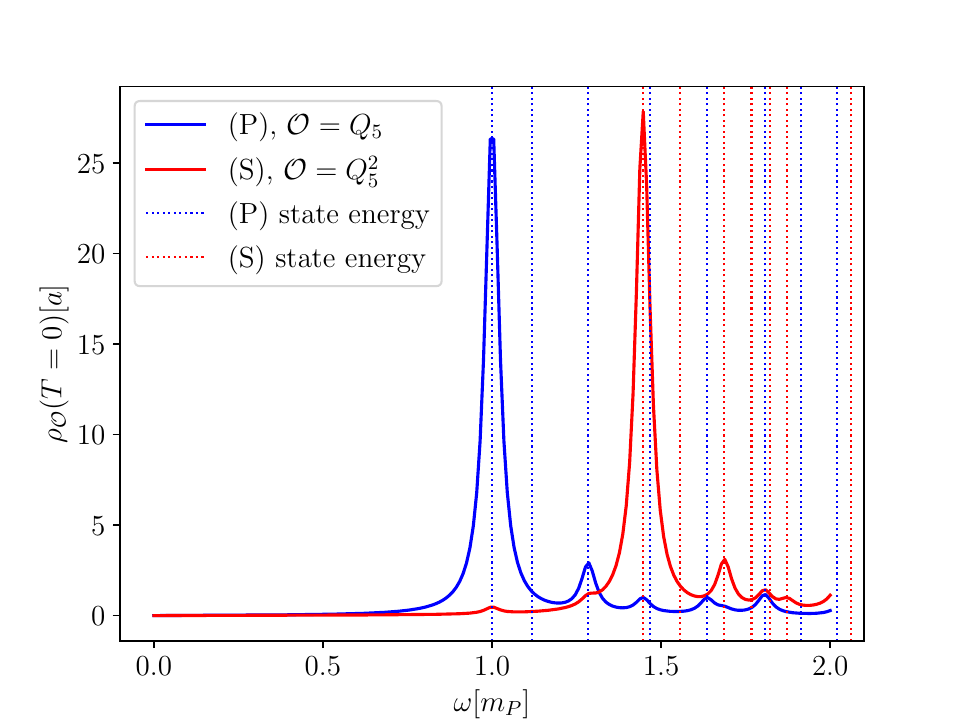}
    \caption{Spectral function in the vacuum in the pseudoscalar (P) and scalar (S) sectors. Energies of the first few excited states are shown in \newtext{dotted lines with blue color} for negative (P) and \newtext{red color for} positive (S) ${\cal C}$-parity. \newtext{The corresponding energy levels are presented in Table \ref{tab:spectrum}.} The parameters of the system are $m/g=0.5, N=12, ga=0.5, \epsilon a = 0.02$. Energy is measured in the units of the first excited state energy that corresponds to the mass of the pseudoscalar meson $m_P$.}
    \label{fig:vac_spectr_func}
\end{figure}

From the structure of Eq.~\eqref{eq:freq_current_correlator} it is clear that the peaks in the spectral function correspond to the differences in the energies of the Hamiltonian eigenstates, with contributions from different pairs of states weighed by the thermal coefficients $e^{-\beta E_n}$ and the matrix elements of the axial charge operator, $c_{mn}$. In Fig.~\ref{fig:vac_spectr_func} we display the spectral function in the vacuum in the pseudoscalar (with ${\cal O} = Q_5$) and scalar (${\cal O} = Q_5^2$) sectors and show the energies of several low-lying excited states for comparison. We differentiate between the excited states $|\Psi\rangle$ by the sign of the $\langle S_R\rangle_\Psi$, where  $S_R\equiv\otimes_{i=1}^N X_i T^{(1)}$ , with 
$T^{(1)}$ denoting the cyclic translation by one site. In the case of periodic boundary conditions this operator commutes with the Hamiltonian and measures $\cal C$-parity of the state. With open boundary conditions, charge conjugation is no longer an exact symmetry of the Hamiltonian, but the expectation value of this operator can serve as an approximate way to distinguish scalar and pseudoscalar states~\cite{Banuls:2013jaa, Itou:2023img}. \newtext{Namely, we assign a positive $\cal C$-parity to states in the Hamiltonian spectrum $|\psi\rangle$ such that $\langle\psi|S_R|\psi\rangle>0$ and vice versa for negative parity. Energies and parity expectation values for a few low-lying states in the Hamiltonian at $N=12, ga=0.5, m/g=0.5$ are listed in Table \ref{tab:spectrum}.} As is observed in Fig.~\ref{fig:vac_spectr_func}, the peaks in the pesudoscalar spectral function originate mainly from the pseudoscalar states, with the similar matching in the scalar sector. We note that the low amplitude peaks for both operators correspond to excited states with the same quantum numbers of the respective meson, but with a finite momentum. 

\begin{table}[h!]
\centering
\begin{tabular}{|c|c|c||c|c|c||c|c|c|}
\hline
$n$ & $E[1/a]$ & $\langle S_R\rangle$ &
$n$ & $E[1/a]$ & $\langle S_R\rangle$ &
$n$ & $E[1/a]$ & $\langle S_R\rangle$ \\
\hline
0  & 0.0000 &  0.8063 & 5  & 1.1306 & -0.3715 & 10 & 1.3918 & -0.4487 \\
1  & 0.7705 & -0.6196 & 6  & 1.1996 &  0.3136 & 11 & 1.4041 &  0.0808 \\
2  & 0.8615 & -0.6428 & 7  & 1.2605 & -0.1209 & 12 & 1.4430 &  0.0951 \\
3  & 0.9899 & -0.4005 & 8  & 1.3000 &  0.2920 & 13 & 1.4754 & -0.4685 \\
4  & 1.1145 &  0.7293 & 9  & 1.3618 &  0.0945 & 14 & 1.5555 & -0.0997 \\
\hline
\end{tabular}
\caption{\newtext{Low-lying spectrum of the Schwinger model with $m/g=0.5, N=12, ga=0.5$. $n$ denotes the state level, starting from 0 for the vacuum. Energies $E$ have vacuum energy subtracted and are given in the units of inverse lattice spacing $a$. The sign of expectation values of the approximate $\cal C$-parity operator $\langle S_R\rangle$ determines whether the state is shown as a pseudoscalar or scalar in Fig. \ref{fig:vac_spectr_func}.}} \label{tab:spectrum}
\end{table}

\begin{figure}[h!]
    \centering
    \includegraphics[width=\linewidth]{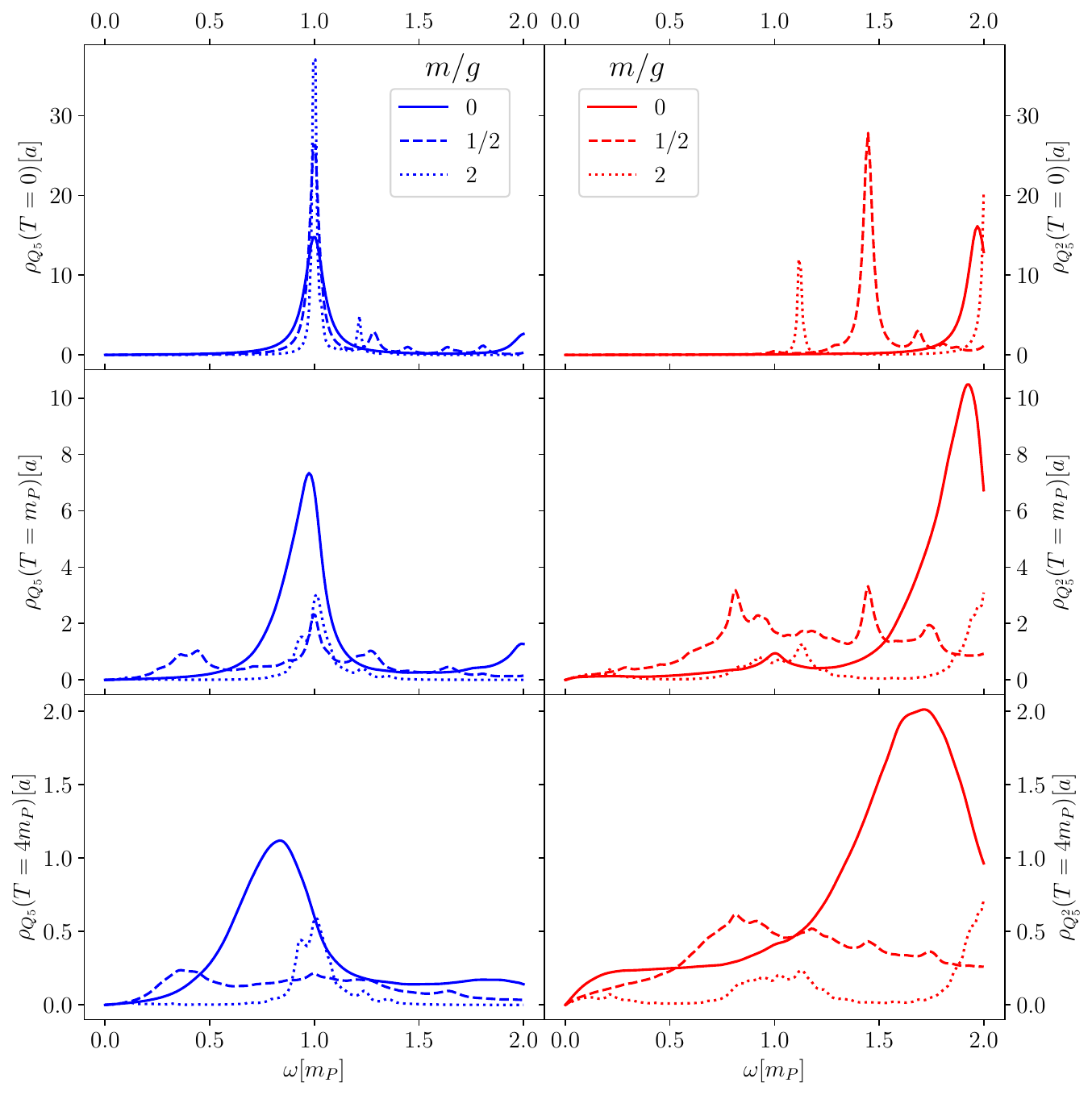}
    \caption{Spectral function at zero and finite temperature for the pseudoscalar and scalar operators, with three values of the fermion mass compared. In all cases, $N=12, ga=0.5, \epsilon a = 0.02$.}
    \label{fig:therm_spectr_func}
\end{figure}

The spectral functions in the vacuum and at finite $T$ are shown in Fig.~\ref{fig:therm_spectr_func} for a variety of $m/g$ and temperature values. In the vacuum, the leading peaks are at the corresponding meson masses, $m_P$ and $m_S$. The fate of these peaks at high temperatures depends on the fermion mass, which we follow to analyze. When $m/g=0$ the theory is that of free bosons, and thus the thermal medium can not \textit{screen} the inter-fermionic potential of the Schwinger boson. Note that in this case, as discussed above, the peak in the spectral function for the scalar state is very close to two times the mass of the pseudoscalar meson. In the opposite limit, when $g/m\to 0$ ($m/g=2$ in the numerical results), the theory approaches a free fermion model, where bound states are not formed, thus e.g. the pseudoscalar peak appears at $2m$. For both of these cases, we observe that the peaks at $T=0$ survive at finite temperature, while no lower frequency mode appears. Note that while for heavy fermions the peak's positions do not shift, for the massless case we observe an overall shift towards lower frequencies with increasing temperature.

In contrast, in the scenario where $m/g=0.5$, the theory is genuinely interacting. Indeed, in this scenario we observe that both peaks in the spectral functions decay as $T$ increases, virtually disappearing once $T/m_P=4$. At the same time, we observe that both the pseudoscalar and the scalar cases develop a low frequency peak.

We can further quantify the connection between the thermal modifications to mesons in this theory to what is observed in the QCD context. In the latter, it is expected that, at finite $T$, the inter-quark potential develops an imaginary component, associated to the suppression (\textit{melting}) of the bound state. At the same time, the real part of the potential should remain mostly unchanged at these temperatures, as a thermal medium does not alter the states intrinsic properties. This behavior has been observed in non-perturbative extractions of the complex heavy-quark potential from (Euclidian) lattice QCD, which find a temperature-dependent imaginary part but a nearly temperature-independent real part, see e.g.~\cite{Larsen:2019zqv,Burnier:2014ssa,Rothkopf:2011db, Tang:2025ypa,Ding:2025fvo,Bazavov:2023dci,Bala:2021fkm,Shi:2021qri,Larsen:2019bwy}.

\begin{figure}
    \centering
    \includegraphics[width=1\linewidth]{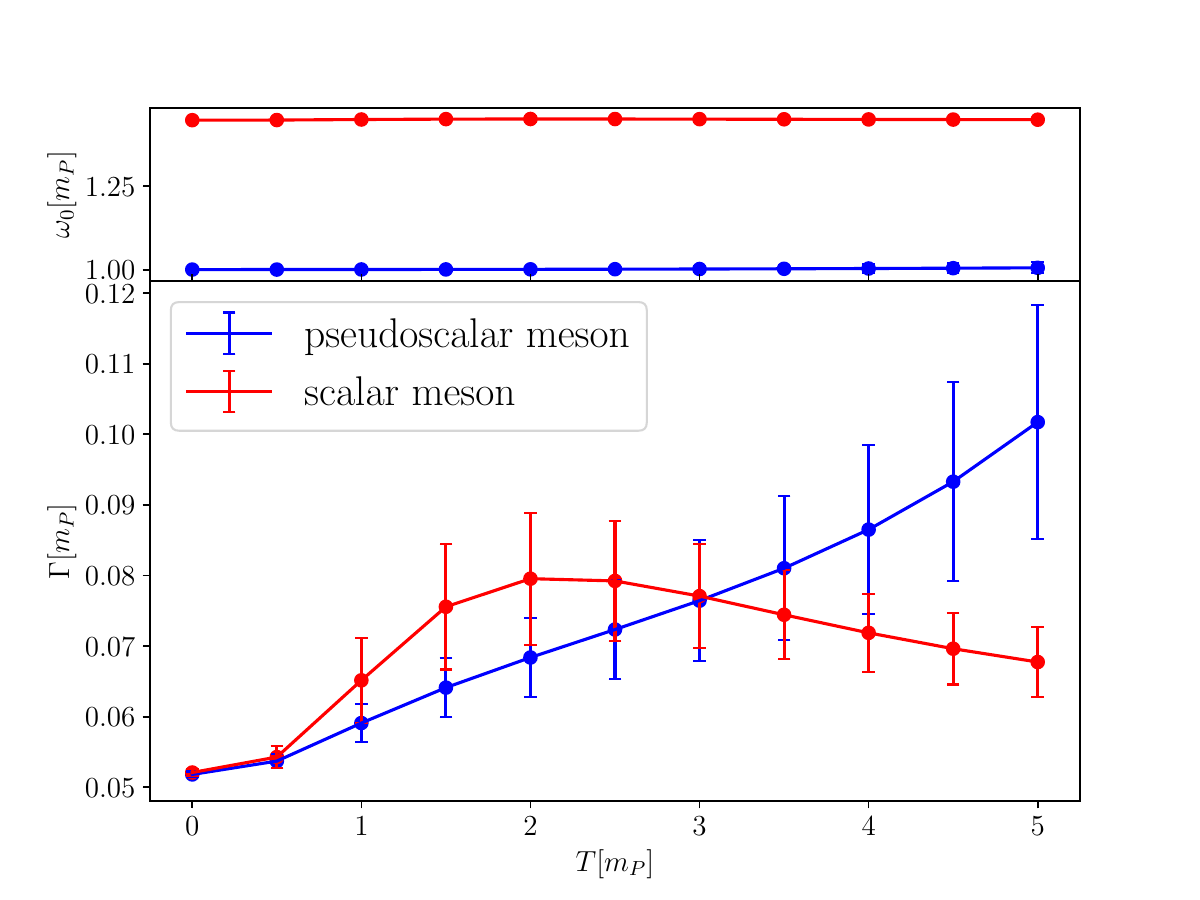}
    \caption{Frequency (upper panel) and width (lower panel) of the pseudoscalar and scalar meson peaks in the corresponding spectral functions, as functions of temperature. Thermal modifications broaden both peaks but leave the meson mass intact. Here we considered the same set conditions as in Fig.~\ref{fig:therm_spectr_func} with $m/g=0.5$.}
    \label{fig:peak_widths}
\end{figure}

In the present context, the previous observations imply that the characteristic frequency of the spectral function peak should not evolve with $T$, while its width should increase with temperature.  Figure~\ref{fig:peak_widths} corroborates this expectation: the upper
panel shows no statistically significant drift of the peak frequency $\omega_{0}$ for either the
scalar or the pseudoscalar state, while the lower panel displays a clear broadening of both peaks with widths $\Gamma$ obtained from Breit--Wigner fits. This behavior provides \textit{ab~initio} evidence that bound--state thermal modification is governed by the emergence of a finite imaginary part of the inter--quark potential, without an appreciable change in its real part. Interestingly, the pseudoscalar width grows faster than the scalar one. While this trend is opposite to a strong--coupling based intuition, where one expects the size of the scalar state to be larger than the pseudoscalar, the large error bars at the highest temperatures preclude a definite conclusion; simulations on larger lattices will be required to better understand the detailed evolution with temperature. \newtext{Finally, we note that the magnitude of the spectral function depends strongly on the choice of an operator in Eq.(\ref{eq:2pt}) and does not have a direct physical meaning. In particular, it cannot be compared to any experimental observables in QCD. For the same reason, no physical conclusions are to be drawn from a comparison between the heights of the peaks in the spectral function for the different $m/g$ and $T$.} 

\vspace{0.3cm}

\emph{\textbf{Energy-Energy correlators.}}
High energy collider experiments allow to probe the small scale properties of 
quantum field theories by measuring the final state of scattering experiments through the use of energy flow detectors. On the theory side, the role played by calorimeter detectors is captured by the notion of asymptotic detector observables. The properties of the theory are encoded in the many-point correlation functions of such objects, see e.g.~\cite{Hofman:2008ar,Sveshnikov:1995vi,Basham:1978bw,Tkachov:1995kk}, which have seen a revamped development in recent years~\cite{Moult:2025nhu}.

In the context of real-time simulations, one can study the properties of these asymptotic operators beyond the limitations of real detectors and traditional theoretical methods, which either rely on perturbation theory or that the theory has sufficient underlying symmetries. Here, following~\cite{Barata:2024apg}, we consider the one and two point energy correlation functions: 
\begin{align}
    C_{1,\pm}(\beta) &= \Tr\,[\rho_\beta \,{\cal O}^\dagger \,{\cal C}_\pm\, {\cal O}]\,, \nn 
    C_2(\beta) &= \Tr\,[\rho_\beta \,{\cal O}^\dagger \,{\cal C}_+\, {\cal C}_-\, {\cal O}]\,,
\end{align}
where $\pm$ indicates the position of the detector/flow operator $\mathcal{C}$ on the left or right edges of the system, and the $\cal O$ operator generates a local excitation on top of the equilibrium thermal state. The flow operators are obtained by light-transforming locally conserved currents~\cite{Hofman:2008ar}; in the case of the energy and electric charge they take the simplified form in $1+1$-d:
\begin{align}
   {\cal Q_\pm}&\equiv \pm \int_0^\infty dt\,J^1(t, x\rightarrow\pm\infty) \,  , \nn 
    {\cal E_\pm}&\equiv \pm\int_0^\infty dt\,T^{01}(t, x\rightarrow\pm\infty)\, .
\end{align}
Since in the continuum and infinite volume limits the states of the theory are charge neutral, here we restrict the discussion to energy correlators. For this, the off-diagonal component of the energy momentum tensor can be written as~\cite{Janik:2025bbz}
\begin{align}
    T^{01}(t,x_n)&= e^{iHt} T^{01}(t=0,x_n)e^{-iHt}\nn
    &\hspace{-1.5 cm}= e^{iHt} \left(\frac{Y_n Z_{n+1} X_{n+2} - X_n Z_{n+1} Y_{n+2}}{2a^2}\right) e^{-iHt} \,,
\end{align}
after performing the Jordan-Wigner transformation.

To excite the system, we consider a (scalar) quench proportional to the local kinetic energy term at the center of the lattice; the excitation operator reads
\begin{equation}
    {\cal O} = X_{N/2}X_{N/2+1} + Y_{N/2}Y_{N/2+1} \, .
\end{equation}
In the strong coupling regime this corresponds to the creation operator of a spatially localized meson state, with $X$ and $Y$ operators flipping the spins at the corresponding sites. This particular combination of $X$ and $Y$ has the virtue of conserving the global charge, i.e. $[{\cal O}, \sum_n Z_n ] = 0$, and thus corresponds to a gauge-invariant operator in the Schwinger model language. We note that even though one could, for example, consider a vector excitation, in a $1+1$-d model such a quench would not introduce any new element, unlike what occurs in higher dimensions where correlation functions acquire non-trivial angular modulations. We also consider the linear rescaling of the scalar excitation
\begin{equation}
    {\cal O}_\alpha \equiv (1-\alpha)\, I + \alpha\, {\cal O} \, ,
\end{equation}
which allows to adjust the amount of energy injected into the system by changing the parameter $\alpha$. In what follows, we shall refer to the choice $\alpha=1$ as \textit{strong quench} and we reserve the term \textit{weak quench} for the choice $\alpha = 0.17$, which is empirically found to deposit the same amount of energy in the vacuum, as the strong quench does in the thermal state with $T=2m_P$.

\begin{figure}
    \centering
    \includegraphics[width=1\linewidth]{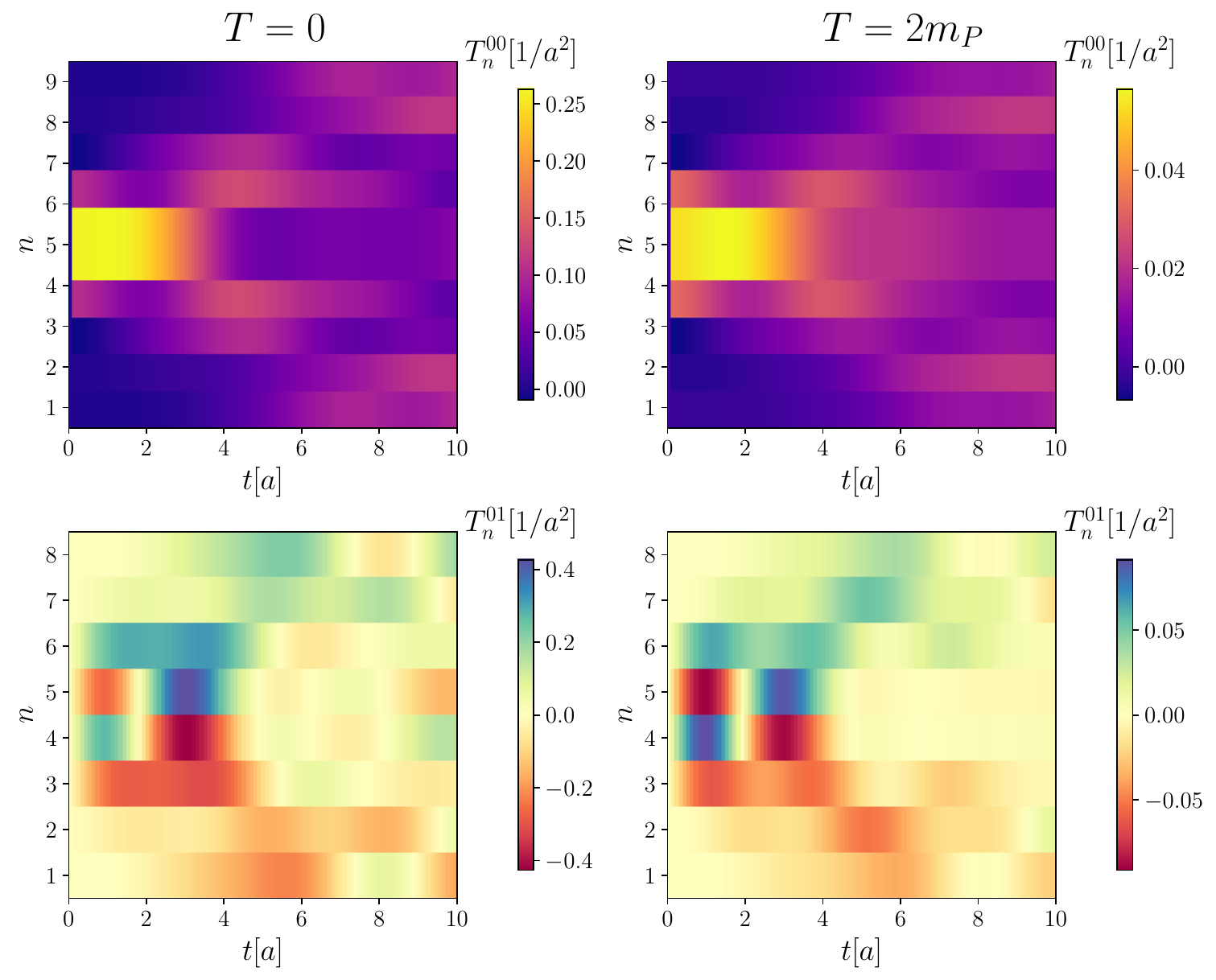}
    \caption{Energy (top panel) and momentum (bottom panel) densities as functions of space and time for a strong quench in the vacuum ($T=0$, left panel) and at finite temperature ($T=2m_P$, right panel).}
    \label{fig:energy-1-point}
\end{figure}

In Fig.~\ref{fig:energy-1-point} we show the spacetime evolution of the energy ($T^{00}$) and momentum ($T^{01}$) densities, after the initial quench, at zero (left) and finite $T$ (right). For both quantities, after the initial quench they spread along a light-cone trajectory, until reaching the lattice's edge, after which a boundary induced oscillatory pattern ensues. Despite the small size of the lattice, we observe that the finite and zero $T$ results show a similar light-cone structure, although slightly smeared at finite temperature. Throughout this section we present the results for a system with $N=10$, $ga = 0.5$ and $m/g=0.5$. We do not show the data for $t>10$, where edge effects 
become prevalent.

To capture the evolution of correlations in the energy flows, we compute the \textit{time differential} energy-energy correlators 
\begin{align}
    D(t,t')\equiv{\cal N}\,\Tr\,[\rho_\beta \,{\cal O}^\dagger \,T^{01}(t,+)\, (-T^{01})(t',-)\, {\cal O}]\,, \label{eq:time_dif_EE}
\end{align}
whose time integral corresponds to $\mathcal{C}_2$, ${\cal N}\equiv [\Tr(\rho_\beta {\cal O}^\dagger{\cal O})]^{-1}$ is the normalization prefactor, and  $+$ and $-$ in the arguments of $T^{01}$ correspond to the physical locations at the right and left edges of the system, respectively.

\begin{figure}
    \centering
    \includegraphics[width=1\linewidth]{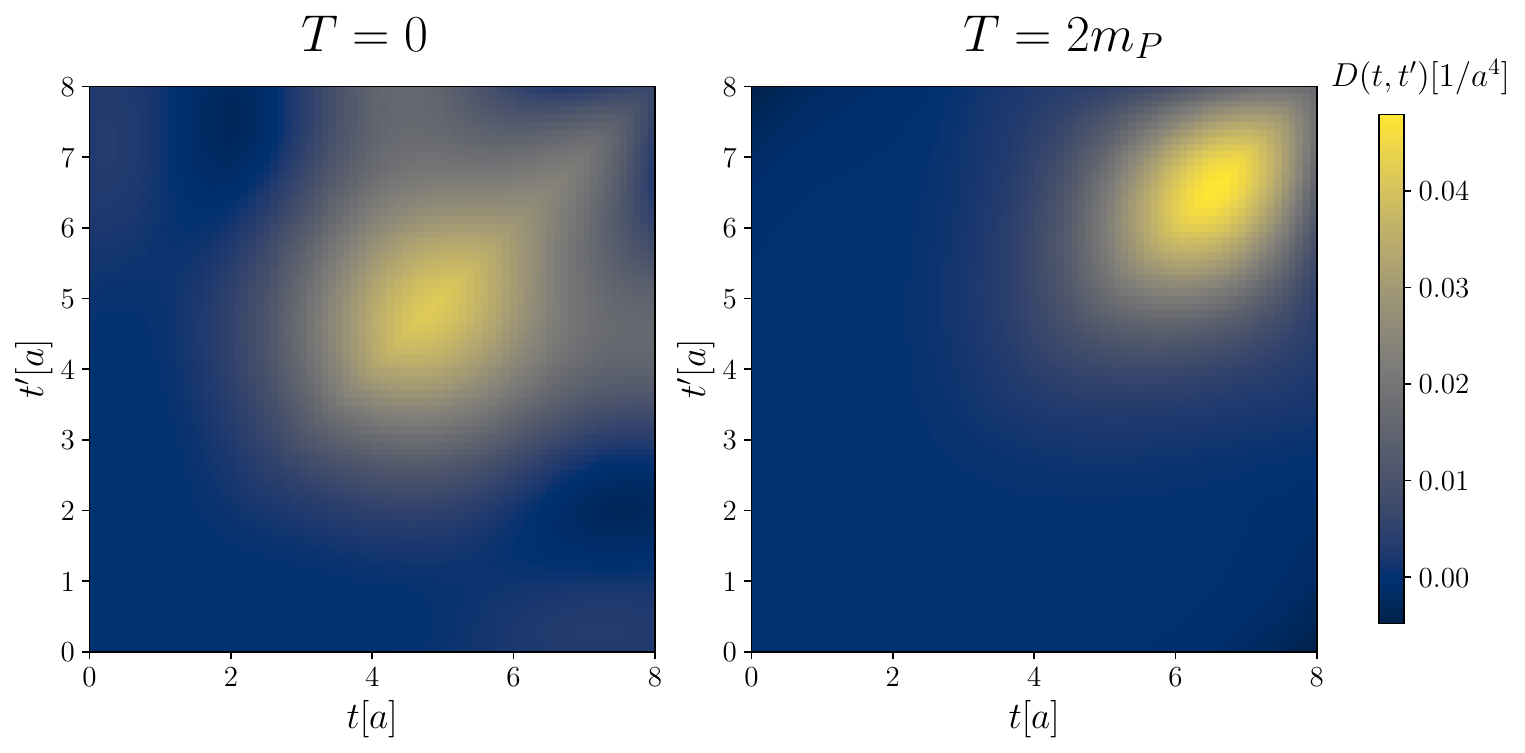}
    \caption{Energy-energy correlator for a strong quench in the vacuum (left panel) and at finite $T$ (right panel).}
    \label{fig:2_pt_en}
\end{figure}

In Fig.~\ref{fig:2_pt_en} we present the results for the time differential correlators. Correlations remain small at early times, when the excitations have not yet reached the detectors. In the subsequent times, we observe a peak along the line $t=t'$ centered around the time the light-cone reaches the detector, and decreasing once the signal propagates past it. We only show data up to $t,t'=8a$, before the edge effect is visible. Comparing the zero and finite temperature results, we observe that the vacuum energy-energy correlator is rather sensitive to out of time correlations, while the finite $T$ scenario the build up of correlations is more peaked, occurring at later times and more concentrated along $t=t'$. 

\begin{figure}
    \centering
    \includegraphics[width=1\linewidth]{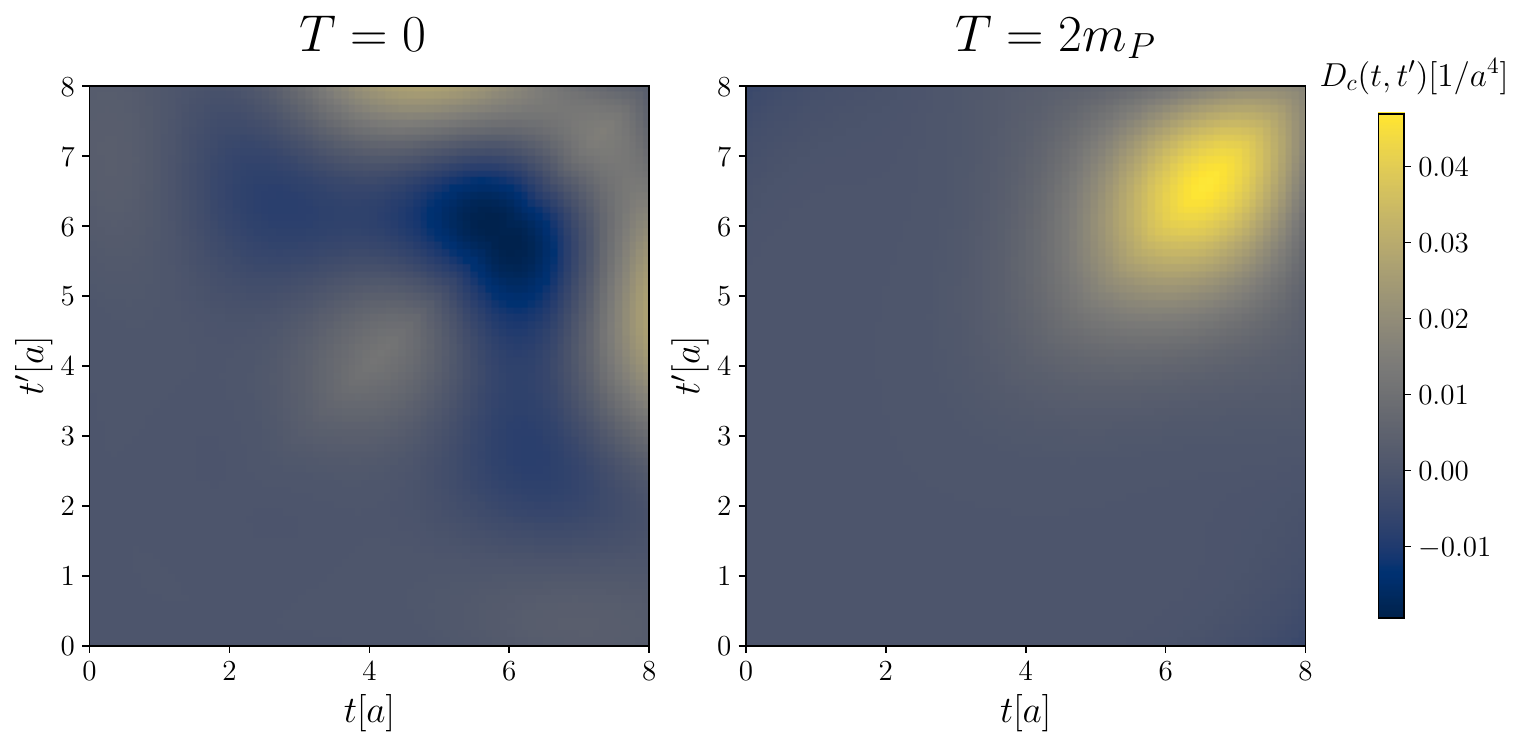}
    \caption{Connected energy-energy correlator for a strong quench in the vacuum (left panel) and at finite $T$ (right panel).}
    \label{fig:conn_2_pt_en}
\end{figure}

We further quantify how non-trivial correlations build up after the quench by computing the connected two point correlator 
\begin{align}
    D_c(t,t')=D(t,t')-D_+(t)D_-(t')\, ,
\end{align}
where $D_\pm(t)$ denotes the one point time differential correlator along one of the lattice's edges. If this quantity is vanishing at all points, or in a non-zero measure set, then one would conclude that the energy-energy correlator factorizes: $\langle  \mathcal{E}_+\mathcal{E}_-\rangle = \langle\mathcal{E}_+ \rangle \langle \mathcal{E}_-\rangle $. In Fig.~\ref{fig:conn_2_pt_en} we show the evolution of the connected two point correlator at zero and finite $T$. Remarkably the connected correlator at finite $T$ exhibits larger values (around two to four times larger at the peak), compared to the vacuum case. This seems to suggest that the inclusion of the finite $T$ bulk results in the build-up of non-trivial correlations in the state, which are reflected in the outgoing momentum flows. The small values observed in the vacuum are perhaps not surprising due to the $1+1$-d nature of the theory and the low level of entanglement generated by the quench used. A better quantification of these observations requires a larger system size lattice study, which we leave for future work.
 
\begin{figure}
    \centering
    \includegraphics[width=1\linewidth]{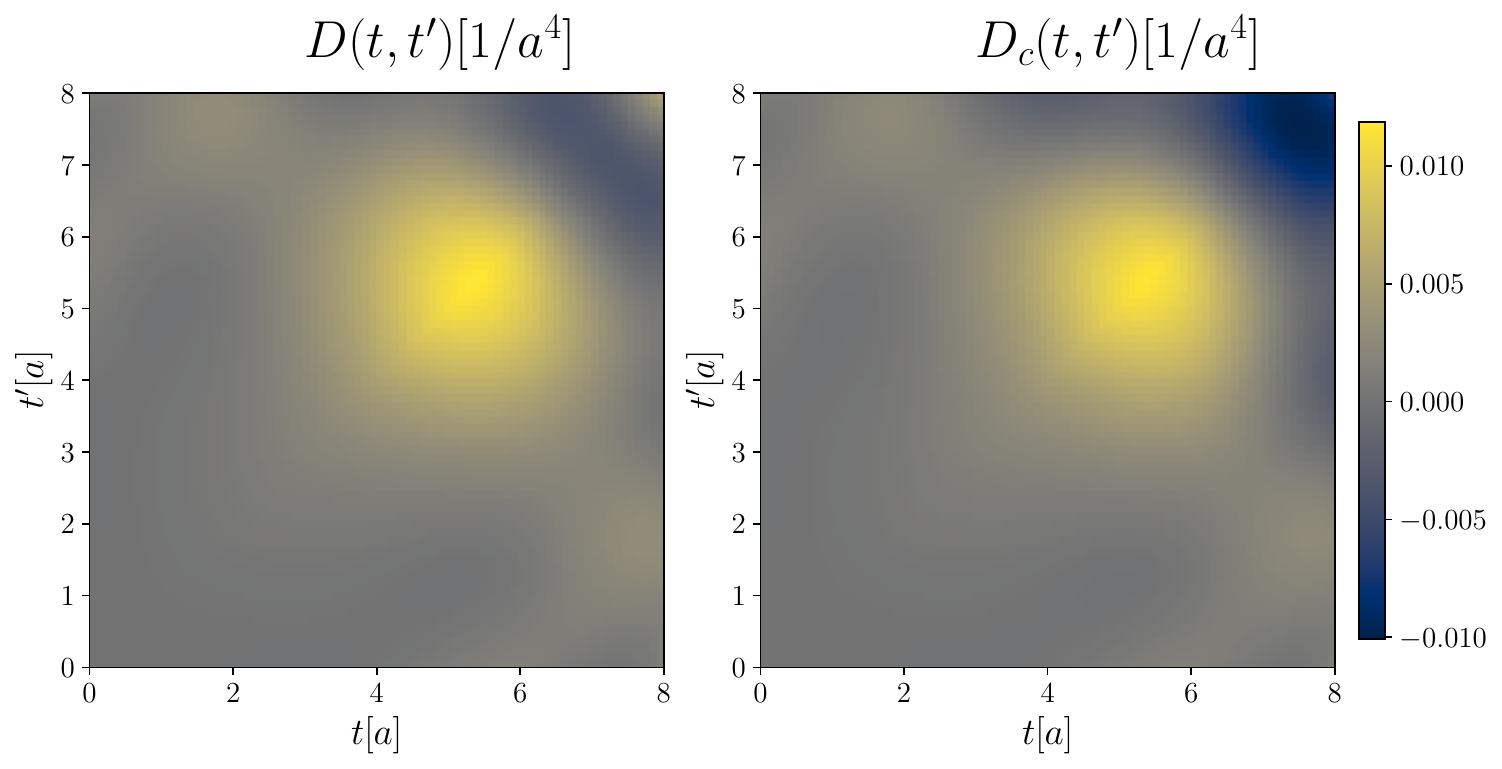}
    \caption{Left panel: energy-energy correlator for a weak quench in the vacuum. Right panel: corresponding connected energy-energy correlator for a weak quench in the vacuum.}
    \label{fig:weak_quench}
\end{figure}

Finally, in Fig.~\ref{fig:weak_quench} we consider a weak quench which allows to compare the vacuum and finite temperature cases when the same energy is deposited into the system. 
We note that although $D\approx D_c$ in this case, the single point correlators are numerically rather small. Still, the structure observed for the connected correlator matches more closely what is seen for the strong quench at finite temperature, although the magnitudes are different.

\newtext{We stress that the quench that we used has no particular meaning from the viewpoint of QCD and we do not expect that it corresponds to any real collider process. For this reason added to the fact that we work in a toy model, the numerical values of the energy correlators bear no meaning for the (3+1)-dimensional QCD. Furthermore, the real particle detectors do not provide sufficient resolution to study time-differential energy correlators, introduced in Eq. (\ref{eq:time_dif_EE}). Nevertheless, out results help frame the general expectations about temperature dependence and factorization properties of energy correlators, that can be measured in real processes.}

\vspace{0.3cm}

\emph{\textbf{Conclusion.}}
In this work, we have studied the properties of the Schwinger model at finite temperature. In particular, we have focused in extracting the spectral functions of the lowest lying meson excitations and the correlations between asymptotic energy flow operators, which have clear connections to QCD in the context of heavy ion collisions. To this end, we have employed real-time lattice simulations, which allow to explore the real-time properties of the model underlying these observables without encountering sign problems.

Analogously to what is observed in heavy ion collisions, we have found that when placing mesons in a finite temperature plasma, 
the corresponding spectral widths increase with $T$. This indicates the enhancement of imaginary part of the respective inter-fermionic potential, responsible for the suppression of bound states as observed in QCD. Further, in the weak and strong coupling limits, where the theory is effectively free, we observe that the mesons' spectral functions do not develop a low frequency peak, as what occurs in the interacting theory. Our result correspond to a first demonstration of thermal broadening of mesons, from real-time simulations, in a gauge theory.

We have further studied how correlations in the momentum flows build up after a sudden quench, also in direct analogy to what occurs in high energy scattering experiments. We have found that at finite $T$ the connected correlator is enhanced compared to the vacuum case, indicating the build-up of non-trivial correlations in the hot bulk; this matches the expectations gained from experimental results in high energy nuclear scattering experiments, see e.g.~\cite{CMS-PAS-HIN-23-004,talk_Anjali}.

Looking forward, it would interesting to go beyond the small systems sizes here considered, either using quantum simulation or tensor network techniques, following the recent interest in exploring the finite $T$ dynamics of gauge theories using these methods, see e.g.~\cite{Angelides:2025hjt,Du:2023ewh,Ikeda:2024rzv,Qian:2024xnr,Saito:2014bda,Pedersen:2023asd, Florio:2025hoc, Barata:2023clv,Kuramashi:2018mmi}. On the modeling side, it would be interesting to extend the current considerations to multiflavor models~\cite{Sadzikowski:1996em, Banuls:2016gid,Hetrick:1995wq} or including a topological term~\cite{Itou:2024psm, Dempsey:2023gib} to the action. Such modifications would allow for the formation of closer analogs to the QCD mesons states, due to mass splitting in the fermions, or provide a new way to explore other aspects of bound state modification at fixed $T$. It would also be valuable to understand how the meson observables considered here, which are frequently used in QCD, relate to QIS inspired measures, as explored in e.g.~\cite{Banuls:2022iwk}. Another interesting opportunity lies in relating the observations made for the finite $T$ connected energy-energy correlator to what has been observed in heavy ion collisions, where both classical and quantum correlations play an important role, see discussion in e.g.~\cite{Barata:2025fzd, Yang:2023dwc}.

\section*{Acknowledgments}
We are grateful to Mari Carmen Bañuls, Adrien Florio, Enrique Rico, and Wenyang Qian for helpful discussions. This material is based upon work supported by the U.S. Department of Energy, Office of Science, Office of Nuclear Physics under contract number DE-SC0012704, and Laboratory Directed Research and Development funds (project No. 25-033B) from Brookhaven Science Associates.

\bibliographystyle{JHEP-2modlong.bst}

\bibliography{Lib.bib}

\end{document}